\documentclass[conference,9pt]{IEEEtran}
\IEEEoverridecommandlockouts

\usepackage{cite}
\usepackage{amsmath,amssymb,amsfonts}
\usepackage{algorithmic}
\usepackage{graphicx}
\usepackage{textcomp}
\usepackage{xcolor}

\usepackage{comment}
\usepackage{cleveref}
\usepackage{multirow}
\usepackage{makecell}
\usepackage{adjustbox}
\usepackage{cite}
\usepackage{url}
\usepackage{titlesec}

\usepackage{etoolbox}
\newbool{blind}
\boolfalse{blind} 

\usepackage{pifont}
\newcommand{\cmark}{\ding{51}}%

\newcommand{\asrvocab}{ASR}%
\newcommand{\llmvocab}{Qwen}%

\newcommand{\argmax}[1]{\underset{#1}{\operatorname{argmax}}}

\newcommand{\eos}{\ensuremath{\texttt{EOS}}}
\newcommand{\bos}{\ensuremath{\texttt{BOS}}}

\newcommand\numberthis{\addtocounter{equation}{1}\tag{\theequation}}

\graphicspath{{../figures/}}

\def\BibTeX{{\rm B\kern-.05em{\sc i\kern-.025em b}\kern-.08em
    T\kern-.1667em\lower.7ex\hbox{E}\kern-.125emX}}

\begin{document}

\title{LLMs and Speech: Integration vs. Combination\\
}

\ifbool{blind}{
  \author{\IEEEauthorblockN{Anonymous Authors}
\IEEEauthorblockA{\textit{Anonymous Affiliation} \\
City, Country \\
anonymous@affiliation.com}
  }
}{

  \author{\IEEEauthorblockN{Robin Schmitt, Albert Zeyer, Mohammad Zeineldeen, Ralf Schlüter, Hermann Ney}
\IEEEauthorblockA{\textit{Machine Learning and Human Language Technology Group, Faculty of Computer Science,} \\
\textit{RWTH Aachen University, Aachen, Germany} \\
\textit{AppTek GmbH, Aachen Germany}\\
\{schmitt,zeyer,zeineldeen,schlueter,ney\}@ml.rwth-aachen.de}
}
}

\maketitle

\begin{abstract}
In this work, we study different approaches to utilize
large language models (LLMs) for automatic speech recognition (ASR).
Specifically, we compare the tight integration of an acoustic model (AM)
with the LLM ("speech LLM")
to the traditional way of combining AM and LLM via shallow fusion
and provide ablations on the effect of different label units and LLM sizes.
For tight integration, we further examine the effect of
attention interfaces,
encoder downsampling,
and length normalization.
Furthermore, we investigate joint recognition with a CTC model to mitigate hallucinations of
speech LLMs and present effective optimizations.
We train and evaluate on LibriSpeech and Loquacious and additionally evaluate on the HuggingFace
ASR leaderboard.
Across model sizes, we find that shallow fusion consistently outperforms
tight integration of AM and LLM on in-domain data,
highlighting the importance of strong shallow-fusion baselines when evaluating speech LLMs for ASR.
On the more heterogeneous HuggingFace ASR leaderboard, however, the integrated prefix LLM 
achieves lower average WER than shallow fusion, with gains concentrated on out-of-domain corpora.
\end{abstract}


\section{Introduction \& Related Work}
Traditionally, speech recognition systems consist of an acoustic model (AM) and a separate language model (LM),
which allows utilizing large text-only corpora in addition to the typically smaller speech corpora.
Multiple approaches for the combination of AMs and LMs have been proposed over the years,
such as shallow fusion, deep fusion \cite{gulcehre2015deepfusion}, cold fusion \cite{sriram2017coldfusion},
internal language model correction \cite{zeineldeen2021:ilm},
and sequence discriminative training \cite{vesely2013sequence}.
In recent years, many techniques utilizing large language models (LLMs) for automatic speech recognition (ASR) 
have been proposed. 

Delayed fusion \cite{hori2025delayedfusion} allows for efficient
single-pass combination of AM and LLM scores and alleviates the problem of label unit mismatch.
$N$-best rescoring \cite{yang2023taskactivatingprompting} can be seen as a special case of
delayed fusion where the LLM score is only used for the final ASR hypotheses.
Error correction \cite{yang2023taskactivatingprompting,chen2023hyporadise} uses an LLM
to generate a corrected transcript from an ASR hypothesis list.
Transducer-LLaMA \cite{denqi2025transducerllama} is a transducer-based ASR system
which uses a pre-trained LLM as the prediction network.
The most common approach, commonly referred to as \textit{speech LLM} (SLLM), is to train an
LLM to also accept acoustic information from an audio encoder as input.
While earlier works discretized the encoder output and extended the
LLM vocabulary to include audio tokens \cite{rubenstein2023audiopalm,zhang2023speechgpt}, 
current approaches directly use the continuous encoder output.
Different interfaces between encoder and LLM have been proposed.
Most works \cite{tang2024salmonn,bai2024seedasr,xu2025fireredasr,abouelenin2025phi4mini,saon2025granite-speech,shi2026qwen3asr}
prepend the encoder output to the LLM input and use self-attention to attend over the combined sequence.
The self-attention on the prefix part can be causal or bidirectional \cite{gupta2024decoder-only}, commonly
referred to as causal decoder-only and prefix LLM \cite{wang20222prefix-lm}, respectively.
Here, we refer to both as prefix LLMs (PLLM) or, more generally, as prefix LMs (PLM), since the term decoder-only
might be misleading given that we use a non-trivial encoder.
Prefix LMs can use interleaved audio-text inputs to enable streaming \cite{jia2025chunkedllm}.
\cite{kong2024audio-flamingo,yu2023llm-cross-att} add gated cross-attention layers to the LLM decoder,
effectively turning it into an attention-based encoder-decoder (AED) model.

While most works on speech LLMs investigate a large set of downstream tasks to
effectively utilize the broad range of capabilities of the LLM, few focus
specifically on ASR performance \cite{bai2024seedasr,xu2025fireredasr,shi2026qwen3asr}.
What is largely missing from existing literature is a consistent comparison between
speech LLMs and classical ASR systems
under comparable conditions to allow for a comparison of the different approaches
to utilize LLMs for ASR,
rather than evaluating the differences in training data exploited.
To the best of our knowledge, \cite{saon2025granite-speech} and
\cite{xu2025fireredasr} are the only exceptions.
\cite{saon2025granite-speech} compare their model to greedy decoding of their
baseline CTC model which is unfair because greedy CTC lacks label context.
\cite{xu2025fireredasr} compare their model to a baseline AED model but do not
state whether the training time is consistent between the two models.
In other works, different training datasets are used for the baseline ASR model
and the speech LLM.
Furthermore, we are not aware of any work systematically comparing
the tight integration of AM and LLM versus the traditional way of combining AM and LLM via shallow fusion.
In this work, we investigate whether using pre-trained text-based LLMs is beneficial for ASR performance
and, if so, how to best utilize them.
Our contributions are as follows:
\begin{itemize}
    \item We compare tight integration of acoustic encoder and LLM to shallow fusion of CTC and LLM scores.
    \item We compare speech LLMs to baseline ASR systems which were only trained on the paired ASR data.
    \item We investigate joint CTC recognition to mitigate hallucinations of speech LLMs
    and present novel optimizations for this joint recognition.
    \item We conduct ablations on the effect of LLM sizes, label units, 
    attention interfaces, encoder downsampling, fine-tuning strategies, and 
    recognition settings for speech LLMs.
\end{itemize}

To our knowledge, this is the first work to systematically compare the different 
approaches to utilizing LLMs for ASR under comparable conditions.

\section{Shallow Fusion Baseline}

\paragraph{Acoustic Encoder.}
All our models utilize a Conformer-based 
encoder \cite{gulati2020conformer}.
The input of 10ms log mel features is first processed by
a convolutional front-end with a subsampling factor of 6.
This sequence is further processed by a Conformer to produce the encoder output
\begin{equation*}
    h_1^T = \operatorname{Encoder}(x_1^{T'}),
\end{equation*}
where $T = \lceil T' / 6 \rceil$ and $T'$ is the length of the input sequence.

\paragraph{Acoustic CTC output.}
On top of the encoder,
we have a linear projection to the label vocabulary dimension including blank,
and we apply CTC \cite{graves2006ctc} (for training and optionally also for recognition):
\begin{align*}
    p_{\textrm{CTC}}(y_t \mid h_t) &= \operatorname{softmax}(\operatorname{Linear}(h_t)) [y_t], \\
    p_{\textrm{CTC}}(a_1^S \mid h_1^T) &= \sum_{y_1^T \in \mathcal{B}^{-1}(a_1^S)} \prod_{t=1}^T p_{\textrm{CTC}}(y_t \mid h_t),
\end{align*}
where $\mathcal{B}$ is the CTC collapsing function which removes blanks and repeated labels.

\paragraph{Language model.}
We use a standalone Transformer-based language model
\begin{align*}
    p_{\textrm{LM}}(a_1^{S}) &= \prod_{s=1}^{S} p_{\textrm{LM}}(a_s \mid a_0^{s-1})
\end{align*}
where $a_0$ denotes the \bos{} token and $a_{S}$ denotes the \eos{} token.
We use the perplexity metric to evaluate our language models:
\begin{align*}
    \operatorname{PPL}(p_{\textrm{LM}}, a_1^{S}) &= \exp\left( -\frac{1}{S} \sum_{s=1}^{S} \log p_{\textrm{LM}}(a_s \mid a_0^{s-1}) \right) . 
\end{align*}

\subsection{Training}

We train the acoustic model using the CTC loss \cite{graves2006ctc}
$$ L_{\textrm{CTC}} = -\log p_{\textrm{CTC}}(a_1^S \mid h_1^T) . $$
The language model is trained separately using the standard cross-entropy loss on the text data
$$ L_{\textrm{LM}} = -\log p_{\textrm{LM}}(a_1^S) . $$
We optionally use a pre-trained LLM as a starting point and fine-tune it on the transcription data.

\subsection{Recognition}

For recognition, we use time-synchronous beam search \cite{prabhavalkar2023endtoend} to
find the label sequence $\hat{a}_1^{\hat{S}}$ with length $\hat{S}$ that maximizes the 
log-probability of the combined CTC and LM scores:

\begin{align*}
\hat{S}, \hat{a}_1^{\hat{S}} &= \argmax{S, a_1^S} 
\left \{ 
    \log p_{\textrm{CTC}}(a_1^S \mid h_1^T) + \lambda \log p_{\textrm{LM}}(a_1^S)
\right \},   \\[1em]
\end{align*}
where $\lambda$ is a tunable scale.
 
In case of matching vocabulary, the LM score is used for non-blank, non-repetition frames
and we combine the scores before pruning.
In case of a vocabulary mismatch, we use delayed fusion.
Specifically, we exactly follow the fixed-interval fusion in \cite{hori2025delayedfusion}
with interval $I=1$.
This means that we incorporate the LLM score as soon as a whole word has finished.

\section{Integrated Models}

\paragraph{Encoder and CTC output.}
We use the same encoder and CTC output as in the shallow fusion baseline.

\paragraph{Adapter.}
The adapter can optionally further downsample the encoder output,
and then optionally performs a linear transformation to match the decoder input dimension.
The downsampling is applied directly to the encoder output $h_1^T$.
In the simple case, we concatenate consecutive frames \cite{ma2024slam-asr},
which is the same as applying a 1D convolution with kernel size and stride
equal to the downsampling factor.
We also use a variant of \textit{CTC compression} \cite{gaido2021ctc-compression,wu2023decoder-only,zhou2025cjst}
which dynamically compresses the encoder output by taking the probabilities of the auxiliary
CTC model into account.
%
\begin{align*}
    \hat{y}_t &= \operatorname{argmax}_y p_{\textrm{CTC}}(y \mid h_t), \\
    m_t &= (\hat{y}_t = \hat{y}_{t-1}) && \text{(merge mask)} \\
    &\quad \wedge (p_{\textrm{CTC}}(\hat{y}_t \mid h_t) \ge \tau) \\
    &\quad \wedge (p_{\textrm{CTC}}(\hat{y}_{t-1} \mid h_{t-1}) \ge \tau), \\
    i_t &= \begin{cases}
        1 & \text{if } t = 1 \\
        i_{t-1} & \text{if } m_{t}, t \ge 2 \\
        i_{t-1} + 1 & \text{if } \neg m_{t}, t \ge 2
    \end{cases} \numberthis \label{eq:ctc-compression-merged-index}
\end{align*}
where $\tau$ is a threshold hyperparameter.
Then we can compress the encoder output by pooling over the frames with the same index $i_t$:
\begin{align*}
    \operatorname{compress}(h_1^T)_i &= \operatorname{mean}(\{ h_t \mid i_t = i \}), \quad i \in 1, \ldots, i_T .
\end{align*}

\paragraph{Decoder and attention interface variants.}
The probability of the output label sequence $a_1^S$ given the encoder output $h_1^T$ is defined as
\begin{equation*}
	p_{\textrm{Dec}}(a_1^S \mid h_1^T) = \prod_{s=1}^S p(a_s \mid a_0^{s-1}, h_1^T).
\end{equation*}
At each step $s$, the Transformer-based decoder \cite{vaswani2017transformer} 
autoregressively predicts the next label $a_s$ by attending over the encoder output $h_1^T$.
We define $a_0=\bos{},a_S=\eos{}$ as the beginning- and end-of-sequence token, respectively,
where $\eos{}$ implicitly models the probability of the sequence length.

We investigate models with different attention interfaces for connecting
the encoder and the decoder.
\emph{AED models} \cite{chorowski2015asr-aed,chan2016las,vaswani2017transformer}
use dedicated \emph{cross-attention} modules,
where the queries are derived from the decoder 
and the keys and values are derived from the encoder.
\emph{PLMs} \cite{abouelenin2025phi4mini,saon2025granite-speech,shi2026qwen3asr}
\emph{prepend} the encoder output to the decoder input
and use self-attention to attend over the combined sequence,
i.e., the queries, keys, and values are all derived from the combined sequence.
\emph{Merged attention models} \cite{zhang2025t5gemma2}
can be seen as a mixture of the previous two,
where the queries are derived from the decoder,
but the keys and values are derived from the combined sequence.
Compared to PLMs, merged attention reduces computation
since the feed-forward layers only need
to process the text part of the combined sequence,
while still allowing the use of a pre-trained
LLM decoder without changing the architecture.

\subsection{Training}

We train our model using the standard label-wise cross-entropy (CE) criterion
$$ L_{\textrm{Dec}} = -\log p_{\textrm{Dec}}(a_1^S \mid h_1^T) .$$
We also add CTC \cite{graves2006ctc,hori2017attctc} as a loss
$$ L_{\textrm{CTC}} = -\log p_{\textrm{CTC}}(a_1^S \mid h_1^T) ,$$
which is either just an auxiliary loss for training or also used for joint recognition.

We optionally use a pre-trained acoustic model as a starting point for the encoder parameters,
and a pre-trained LLM as a starting point for the decoder parameters.

\subsection{Recognition}

For recognition, we use label-synchronous beam search
on $p_{\textrm{Dec}}(a_1^S \mid h_1^T)$
and optionally log-linearly combine
with the prefix scores of the CTC output
$p_{\textrm{CTC}}(a_1^S \mid h_1^T)$
as in \cite{hori2017attctc}.

For the CTC output,
we also optionally apply CTC compression using \Cref{eq:ctc-compression-merged-index}
with threshold $\tau$
as
\begin{align*}
    s_{\textrm{CTC}'}(y_i \mid h_1^T) &= \max_{t \mid i_t = i} p_{\textrm{CTC}}(y_t {=} y_i \mid h_t), \quad i \in 1, \ldots, i_T \\
    p_{\textrm{CTC}'}(y_i \mid h_1^T) &= \frac{s_{\textrm{CTC}'}(y_i \mid h_1^T)}{ \sum_{y'} s_{\textrm{CTC}'}(y' \mid h_1^T) } .
\end{align*}

For the CTC output,
we also investigate top-$k$ pruning of the CTC output distribution.
In contrast to pruning during search as e.g.~in histogram pruning \cite{van1996adaptive},
the pruning is computed once per utterance after taking the maximum over the time dimension of the posteriors:
\begin{align*}
s_{\textrm{max}}(y \mid h_1^T) &= \max_{t} p_{\textrm{CTC}}(y_t {=} y \mid h_t), \\
\operatorname{TopK}(s_{\textrm{max}}, k) \subset \mathcal{Y}
&\Leftrightarrow
\begin{cases}
|\operatorname{TopK}(\ldots)| = k, \\
\min\limits_{y \in \operatorname{TopK}} s_{\textrm{max}}(y)
\ge \max\limits_{y \notin \operatorname{TopK}} s_{\textrm{max}}(y)
\end{cases} \\
s_{\textrm{topk}}(y_t \mid h_t) &= \begin{cases}
p_{\textrm{CTC}'}(y_t \mid h_t) & y \in \operatorname{TopK}(\ldots) \\
0 & \text{otherwise}
\end{cases} \\
p_{\textrm{CTC}^*}(y_t \mid h_t) &=
\frac{s_{\textrm{topk}}(y_t \mid h_t)}{\sum_{y'} s_{\textrm{topk}}(y' \mid h_t)} .
\end{align*}
Intuitively, this only keeps labels that get a high CTC probability in at least one frame.

When not using the CTC output,
we optionally use length normalization by dividing the step-wise 
log-probability $p_{\textrm{Dec}}(a_1^{\bar{S}} \mid h_1^T)$
by the intermediate sequence length $\bar{S}$.
We apply length normalization after the $\operatorname{TopK}$
operation, which means that it does not have an effect
when using beam size 1.

\section{Experiments}

\begin{table}
	\centering
	\caption{
        \textbf{Word-level perplexities} of the used LMs on Loquacious dev and test,
        comparing the base (pre-trained) LLMs to their fine-tuned counterparts.
        Fine-tuned models are fine-tuned for 2 epochs on the Loquacious
        transcripts. Our own Loquacious LM is trained from scratch on the same data.
    }
	\label{tab:lm-ppls}
	\setlength{\tabcolsep}{3pt}
    \begin{adjustbox}{max width=\linewidth}
    \begin{tabular}{|c|c|c|c|r|r|}
        \hline
        \multirow{2}{*}{\makecell{Base \\ LM}} & \multirow{2}{*}{Size} & \multirow{2}{*}{\makecell{Label \\ units}} & \multirow{2}{*}{Type} & \multicolumn{2}{c|}{PPL} \\
        \cline{5-6}
         & & & & dev & test \\
        \hline \hline
        \multirow{8}{*}{Qwen2} & \multirow{3}{*}{0.5B} & \multirow{2}{*}{\llmvocab} & Base & 213 & 206 \\
        \cline{4-6}
         & & & \multirow{2}{*}{Fine-tuned} & 59 & 60 \\
        \cline{3-3} \cline{5-6}
         & & \asrvocab & & 60 & 60 \\
        \cline{2-6}
         & \multirow{3}{*}{1.5B} & \multirow{2}{*}{\llmvocab} & Base & 151 & 146 \\
        \cline{4-6}
         & & & \multirow{2}{*}{Fine-tuned} & 53 & 53 \\
        \cline{3-3} \cline{5-6}
         & & \asrvocab & & 53 & 53 \\
        \cline{2-6}
         & \multirow{2}{*}{7.0B} & \multirow{2}{*}{\llmvocab} & Base & 135 & 131 \\
        \cline{4-6}
         & & & Fine-tuned & \textbf{45} & \textbf{45} \\
        \hline \hline
        Ours & 0.4B & \asrvocab & From scratch & 72 & 73 \\
        \hline
    \end{tabular}
    \end{adjustbox}
\end{table}

For training, we use LibriSpeech 960h \cite{panayotov2015librispeech}
and the \textit{large} split of the Loquacious \cite{parcollet2025loquacious}
corpus, which contains 25K hours of speech data.
We apply on-the-fly speed perturbation and SpecAugment \cite{park2019specaugment}.
When collecting statistics of the datasets, we found that Loquacious and
LibriSpeech only cover 26\% and 19\% of the labels in the Qwen vocabulary, respectively,
which needs to be taken into account when comparing performance with different vocabularies.
\ifbool{blind}{
  We use our own framework
}{
  We use RETURNN \cite{zeyer2018:returnn}
}
with PyTorch \cite{paszke2019pytorch} backend for training and recognition and
we use ESPnet \cite{watanabe2018espnet} to calculate the CTC prefix scores for shallow fusion.
We will release all relevant code upon publication.

We train baseline AED, prefix LM, and merged attention models from scratch.
All baseline models are either trained for 20 epochs on Loquacious 
or for 100 epochs on LibriSpeech.
Next, we train SLLMs by using the baseline
AED encoder together with a pre-trained LLM for the decoder as the starting point.
For the pre-trained LLM, we compare using the 0.5B, 1.5B, and 7B versions of Qwen2 \cite{yang2024qwen2}.
For fine-tuning,
the weights of the adapter are always initialized randomly.
For the decoder weights, we experiment with 
full fine-tuning and low rank adaptation (LoRA) \cite{hu2022lora}.
We use rank-stabilized LoRA \cite{kalajdzievski2023stable-lora} 
adapters with a rank of 320 on the query and value projection matrices of the
LLM attention blocks, as is done in \cite{saon2025granite-speech}.
We also train SLLMs using the ASR SPM vocabulary where the embedding
layer is initialized based on the pre-trained LLM embeddings
as proposed in \cite{denqi2025transducerllama}.
The encoder is always fully fine-tuned.
By default, we directly train all weights of the model jointly.
We also experiment with a 2-stage training scheme,
where we first only train the newly initialized weights and then fine-tune the whole model jointly.
The prompt to the LLM has the form: 
\begin{align*}
    &\textbf{USER: } \langle \texttt{encoder output} \rangle \\ 
    &\textbf{Transcribe this speech to text. ASSISTANT: } \langle \texttt{BOS} \rangle
\end{align*}
We experimented with simply concatenating the encoder output with the $\texttt{BOS}$ token as prompt but this led to slightly worse results.
The LLM uses causal attention on the prompt positions. 
Using bidirectional attention instead had no effect on performance in our experiments.

All our baseline ASR models use a 16-layer Conformer encoder (406M params) and 
10K sentencepiece (SPM) subword label units \cite{kudo2018sentencepiece}.
Baseline AED, prefix LM and merged attention models use a 6-layer Transformer decoder (133M params).
When training speech LLMs with the original LLM vocabulary, we add two more randomly initialized Conformer layers on top of our
encoder to better adapt to the label units of the LLM, resulting in 457M params.
For all models, we add an auxiliary CTC loss on top of the encoder output which 
uses the same label units as the decoder (10M params for 10K SPM vocabulary and 156M params 
for Qwen (150K BPE) vocabulary).
For the PLLMs, we also experimented with adding a CTC loss on the prefix positions
of the LLM output but this yielded worse results.

For shallow fusion, we use the same pre-trained LLMs as for the speech LLMs.
Furthermore, for Loquacious, we compare using our own Transformer LM which is trained on the ASR transcriptions.
Our LM has 32 layers with model dimension of 1024 and uses the same
10K SPM vocabulary as our ASR models, resulting in 422M parameters.

To improve the perplexities of the pre-trained LLMs for shallow fusion,
we fine-tune them for 2 epochs on the Loquacious ASR transcripts.
We report the resulting perplexities in \Cref{tab:lm-ppls},
where fine-tuning the base LLMs substantially reduces their perplexity.
We apply full fine-tuning for Qwen2 0.5B and Qwen2 1.5B LLMs
and apply LoRA for the Qwen2 7B LLM to reduce computational costs.
Specifically, for Qwen2 7B, we apply rank-stabilized LoRA with rank of 64 and alpha of 128
to all linear layers except the embedding layer and with a LoRA dropout of 0.1.
As for SLLMs, we also fine-tune the LLMs using the ASR SPM vocabulary.
For LR scheduling, we use a warmup phase of 5\% of total training epochs
followed by a cosine decay to a minimum learning rate of 1e-7.
For the full fine-tuning trainings, we use a peak learning rate of 1e-5
while for the LoRA trainings, we use a peak learning rate of 5e-6.

The training transcripts for Loquacious and LibriSpeech are fully uppercased.
We lowercase them to reduce the vocabulary mismatch
between the ASR transcription data and the LLM pre-training data,
which we found to improve performance.



\subsection{Tight Integration vs. Shallow Fusion}


\begin{table}
	\centering
	\caption{
        Comparison of \textbf{tight integration vs. shallow fusion} of acoustic and 
        language model components.
        Comparing label-synchronous beam search of
        prefix LLM (PLLM) vs.~CTC + LLM shallow fusion.
        PLLM+CTC uses the auxiliary CTC model with the same label units as the PLLM decoder.
        LLM+CTC uses the baseline CTC model with 10K SPM label units.
        PLLM encoder is initialized with the baseline AED encoder
        and PLLM decoder/ standalone LLM with the Qwen2 LLM.
        We fine-tune for 2 epochs on Loquacious.
        7B models use LoRA for fine-tuning the decoder, all other models are fully fine-tuned.
    }
	\label{tab:tight-vs-shallow-wers-loq}
    \begin{adjustbox}{max width=\linewidth}
	\setlength{\tabcolsep}{3pt}
    \begin{tabular}{|c|c|c|c|c|c|c|c|}
        \hline
        \multirow{3}{*}{\makecell{LLM \\ size}} & \multirow{3}{*}{\makecell{LLM \\ Label \\ units}} & \multicolumn{6}{c|}{WER [\%]} \\
        \cline{3-8}
        & & \multicolumn{3}{c|}{dev} & \multicolumn{3}{c|}{test} \\
        \cline{3-8}
        &  & PLLM & \makecell{PLLM \\ +CTC} & \makecell{LLM \\ +CTC} & PLLM & \makecell{PLLM \\ +CTC} & \makecell{LLM \\ +CTC} \\
        \hline\hline
        \multirow{2}{*}{0.5B} & \asrvocab & 5.76 & 5.41 & 5.30 & 6.64 & 5.94 & 5.82 \\
        \cline{2-8}
        & \llmvocab & 6.02 & 5.59 & 5.29 & 6.88 & 6.14 & 5.84 \\
        \cline{1-8}
        \multirow{2}{*}{1.5B} & \asrvocab & \textbf{5.47} & \textbf{5.28} & 5.24 & \textbf{6.03} & \textbf{5.80} & 5.80 \\
        \cline{2-8}
        & \multirow{2}{*}{\llmvocab} & 5.73 & 5.48 & 5.22 & 6.62 & 6.06 & 5.78 \\
        \cline{1-1} \cline{3-8}
        7B &  & 5.54 & 5.52 & \textbf{5.18} & 6.13 & 6.05 & \textbf{5.73} \\
        \hline
    \end{tabular}
    \end{adjustbox}
\end{table}

A key question we wish to answer is whether the recently popular tight integration 
of acoustic encoder and language model in the PLLM leads to better performance 
than the more traditional shallow fusion of CTC and a standalone LLM.
We compare tight integration with shallow fusion in \Cref{tab:tight-vs-shallow-wers-loq}.
Across most LLM sizes and vocabularies, we find that shallow fusion outperforms tight integration,
both with and without joint CTC decoding.
The only exception is the Qwen2 1.5B model with 10K SPM vocabulary, where PLLM+CTC is on par with LLM+CTC
on Loquacious test.

Furthermore, \Cref{tab:tight-vs-shallow-wers-loq} shows that joint PLLM+CTC decoding improves
performance over standalone PLLM decoding. 
Looking at the detailed WERs in \Cref{tab:hallucination-detailed-wers-loq}, we find that the improvement stems 
from a reduction in both deletion and insertion errors, while the number of substitutions slightly increases.
The reduction in deletions is consistent across all model sizes and is the main contributor on the dev set.
The reduction in insertions, on the other hand, is most pronounced on the test set and for smaller models.
The insertions of standalone PLLM decoding are 
caused by \textit{oscillations} \cite{frieske2024asr-hallucinations},
i.e. cases where most of the model output is correct, but with the addition of a repeating n-gram.
Often a single or very few sequences account for the majority of these insertions.
Overall, the improvement from joint CTC decoding diminishes with increasing LLM size.

\begin{table}
	\centering
	\caption{
        Effect of \textbf{joint CTC recognition} on the detailed error counts, i.e.\
        substitutions (S), deletions (D), insertions (I), and total errors ($\Sigma$),
        of prefix LLMs (PLLM) for different sizes of Qwen2 on Loquacious.
        A checkmark denotes joint PLLM+CTC decoding, otherwise standalone PLLM decoding is used.
        Models use the 150K BPE Qwen2 vocabulary:
        they are initialized with the baseline AED encoder and a Qwen2 decoder
        and fine-tuned for 2 epochs on Loquacious,
        using LoRA for the 7B decoder and full fine-tuning otherwise.
    }
	\label{tab:hallucination-detailed-wers-loq}
	\setlength{\tabcolsep}{3pt}
    \begin{tabular}{|c|c|c|c|c|c|c|c|c|c|}
        \hline
        \multirow{3}{*}{\makecell{LLM \\ size}} & \multirow{3}{*}{\makecell{Joint CTC \\ recog.}} & \multicolumn{8}{c|}{WER [\%]} \\
        \cline{3-10}
        & & \multicolumn{4}{c|}{dev} & \multicolumn{4}{c|}{test} \\
        \cline{3-10}
            & & S & D & I & $\Sigma$ & S & D & I & $\Sigma$ \\
        \hline\hline
        \multirow{2}{*}{0.5B} &        & 3.45 & 1.49 & 1.08 & 6.02 & 3.75 & 1.53 & 1.59 & 6.88 \\
        \cline{2-10}
                              & \cmark & 3.56 & 1.03 & 1.01 & 5.59 & 3.81 & 1.17 & 1.16 & 6.14 \\
        \hline
        \multirow{2}{*}{1.5B} &        & 3.38 & 1.32 & 1.02 & 5.73 & 3.71 & 1.42 & 1.49 & 6.62 \\
        \cline{2-10}
                              & \cmark & 3.43 & 1.06 & 0.99 & \textbf{5.48} & 3.74 & 1.19 & 1.13 & 6.06 \\
        \hline
        \multirow{2}{*}{7B}   &        & 3.36 & 1.21 & 0.96 & 5.54 & 3.63 & 1.42 & 1.08 & 6.13 \\
        \cline{2-10}
                              & \cmark & 3.43 & 1.09 & 1.00 & 5.52 & 3.70 & 1.23 & 1.12 & \textbf{6.05} \\
        \hline
    \end{tabular}
\end{table}

Lastly, \Cref{tab:tight-vs-shallow-wers-loq} shows that using a specialized ASR vocabulary
improves PLLM performance significantly compared to the original LLM vocabulary, while it has 
only a small effect on shallow fusion (LLM+CTC) performance.

\subsection{Recognition Settings}

\begin{table}
	\centering
	\caption{
        Effect of \textbf{length normalization} on standalone prefix LLM (PLLM)
        recognition, comparing the 10K SPM ASR vocabulary and the 150K BPE Qwen vocabulary.
        Models are initialized with the baseline AED encoder and the Qwen2 0.5B decoder
        and fine-tuned for 2 epochs on Loquacious.
        Model with Qwen vocabulary uses LoRA for fine-tuning the decoder; model with ASR vocabulary is fully fine-tuned.
        Since we apply length normalization after the $\operatorname{TopK}$ operation, it does
        not have an effect when using beam size 1.
    }
	\label{tab:length-norm-vocab-wers-loq}
	\setlength{\tabcolsep}{3pt}
    \begin{tabular}{|c|c|c|c|c|c|}
        \hline
        \multirow{3}{*}{\makecell{Label \\ units}} & \multirow{3}{*}{\makecell{Beam \\ size}} & \multicolumn{4}{c|}{PLLM WER [\%]} \\
        \cline{3-6}
        & & \multicolumn{2}{c|}{dev} & \multicolumn{2}{c|}{test} \\
        \cline{3-6}
        & & \makecell{w/ length \\ norm.} & \makecell{w/o length \\ norm.} & \makecell{w/ length \\ norm.} & \makecell{w/o length \\ norm.} \\
        \hline\hline
        \multirow{4}{*}{\asrvocab} & 1 & \multicolumn{2}{c|}{5.85} & \multicolumn{2}{c|}{6.49} \\
        \cline{2-6}
        & 4 & \textbf{5.76} & 5.89 & 6.64 & \textbf{6.34} \\
        \cline{2-6}
        & 16 & 5.91 & 5.94 & 6.58 & 6.42 \\
        \cline{2-6}
        & 32 & 5.91 & 6.02 & 6.86 & 6.51 \\
        \hline\hline
        \multirow{4}{*}{\llmvocab} & 1 & \multicolumn{2}{c|}{6.13} & \multicolumn{2}{c|}{7.41} \\
        \cline{2-6}
        & 4 & 6.0 & 6.05 & 6.99 & 6.75 \\
        \cline{2-6}
        & 16 & 6.39 & 6.13 & 7.79 & 6.85 \\
        \cline{2-6}
        & 32 & 7.49 & 6.14 & 7.72 & 6.9 \\
        \hline
    \end{tabular}
\end{table}

\begin{table}
	\centering
	\caption{
        Effect of \textbf{beam size} and \textbf{length normalization} for 
        recognition with prefix LLMs (PLLM) on Loquacious.
        Models are initialized with the baseline AED encoder and the Qwen2 0.5B decoder and
        are fine-tuned for 2 epochs on Loquacious using the 10K SPM ASR vocabulary and
        full fine-tuning.
        Since we apply length normalization after the $\operatorname{TopK}$ operation, it does
        not have an effect when using beam size 1.
    }
	\label{tab:beam-size-wers-loq}
	\setlength{\tabcolsep}{3pt}
    \begin{tabular}{|c|c|c|c|c|c|c|c|c|}
        \hline
        \multirow{4}{*}{\makecell{Beam \\ Size}} & \multicolumn{8}{c|}{WER [\%]} \\
        \cline{2-9}
        & \multicolumn{4}{c|}{dev} & \multicolumn{4}{c|}{test} \\
        \cline{2-9}
            & \makecell{w/ length \\ norm.} & \multicolumn{3}{c|}{\makecell{w/o length \\ norm.}} & \makecell{w/ length \\ norm.} & \multicolumn{3}{c|}{\makecell{w/o length \\ norm.}}  \\
        \cline{2-9}
            & \multicolumn{2}{c|}{PLLM} & \makecell{PLLM \\ +CTC} & \makecell{LLM \\ +CTC} & \multicolumn{2}{c|}{PLLM} & \makecell{PLLM \\ +CTC} & \makecell{LLM \\ +CTC} \\
        \hline\hline
        1 & \multicolumn{2}{c|}{5.85} & 5.51 & 5.75 & \multicolumn{2}{c|}{6.49} & 6.04 & 6.40 \\
        \hline
        4 & 5.76 & 5.89 & 5.41 & 5.36 & 6.64 & 6.34 & 5.94 & 5.91 \\
        \hline
        16 & 5.91 & 5.94 & 5.40 & \textbf{5.30} & 6.58 & 6.42 & 5.93 & 5.84\\
        \hline
        32 & 5.91 & 6.02 & 5.40 & \textbf{5.30} & 6.86 & 6.51 & 5.94 & \textbf{5.83} \\
        \hline
    \end{tabular}
\end{table}

\begin{table}
	\centering
	\caption{
        Effect of \textbf{optimizations for joint CTC recognition} with prefix LLMs.
        All models are initialized with the baseline AED encoder and the Qwen2 0.5B decoder.
        We fine-tune for 1 epoch on Loquacious with the 150K BPE LLM vocabulary using LoRA
        for the decoder.
        WER, real time factor (RTF) and maximum GPU memory usage when decoding Loquacious dev
        using a single NVIDIA H100 GPU.
    }
	\label{tab:joint-ctc-topk-compression-wers-loq}
    \begin{adjustbox}{max width=\linewidth}
	\setlength{\tabcolsep}{3pt}
    \begin{tabular}{|c|c|c|c|c|}
        \hline
        \makecell{Top-$k$\\pruning} & \makecell{Compression \\ threshold} & \makecell{Max. GPU \\ mem. [GB]} & RTF [$10^{-3}$] & WER [\%] \\
        \hline\hline
        None & \multirow{2}{*}{None} & 68.9 & 69.0 & 6.00 \\
        \cline{1-1} \cline{3-5}
        10,000 & & 10.8 & 11.5 & \textbf{5.99} \\
        \cline{1-1} \cline{2-5}
        None & \multirow{2}{*}{0.90} & 47.9 & 29.3 & 6.00 \\
        \cline{1-1} \cline{3-5}
        10,000 & & \textbf{\phantom{0}9.3} & \textbf{\phantom{0}6.7} & \textbf{5.99} \\
        \hline
    \end{tabular}
\end{adjustbox}
\end{table}

We investigate the effect of length normalization and beam size for 
standalone PLLM recognition
in \Cref{tab:length-norm-vocab-wers-loq}.
First, we observe that WER almost always degrades for higher beam
sizes.
Second, the effect of length normalization is inconsistent. 
When using the 10K vocabulary, length normalization improves WER
on Loquacious dev while it degrades WER on Loquacious test.
When using the 150K vocabulary, length normalization degrades WER
on both test sets, especially for larger beam sizes.
We investigated this behavior and found that, with length normalization and beam size 32,
the PLLM decoder strongly hallucinates on a small number of hypotheses,
sometimes producing more than 500 insertions in a single utterance.
Without length normalization and larger beam sizes, the opposite effect occurs: 
the PLLM decoder prefers short sequences, producing more deletions.
For all other tables, whenever we report standalone PLLM results, we enable
length normalization depending on the performance on the dev set. 

We further show the impact of beam size for PLLM+CTC and LLM+CTC decoding in \Cref{tab:beam-size-wers-loq}.
For recognitions involving CTC, we never use length normalization and we observe
that higher beam sizes lead to better performance.

We compare different options for optimizing the joint CTC recognition in \Cref{tab:joint-ctc-topk-compression-wers-loq}.
Using top-$k$ pruning with $k=10,000$ and CTC compression with a threshold of 0.9 
reduces the maximum GPU memory usage from 68.9GB to 9.3GB and the RTF from $69.0 \cdot 10^{-3}$ to $6.7 \cdot 10^{-3}$, 
while maintaining the same WER compared to not using any optimizations.

\subsection{Attention Interfaces: From Scratch vs. Fine-tuning}

\begin{table}
	\centering
	\caption{
        Comparing training \textbf{from scratch vs. fine-tuning} and the 
        effect of \textbf{different attention interfaces}
        on LibriSpeech and Loquacious,
        using the 10K SPM ASR vocabulary.
        From-scratch models are trained for 100 epochs on LibriSpeech and 20 epochs on Loquacious.
        Fine-tuned models are initialized with the corresponding baseline AED encoder and a Qwen2 decoder
        and are fully fine-tuned with the 10K SPM ASR vocabulary.
        We fine-tune for 50 epochs on LibriSpeech and 2 epochs on Loquacious.
        Results use joint CTC decoding.
    }
	\label{tab:wers-lbs-attention-interfaces}
	\setlength{\tabcolsep}{3pt}
    \begin{tabular}{|c|c|c|c|c||c|c|}
        \hline
        \multirow{3}{*}{Training} & \multirow{3}{*}{Model} & \multirow{3}{*}{\makecell{Pre-trained \\ decoder size}} & \multicolumn{4}{c|}{WER [\%]} \\
        \cline{4-7}
         & & & \multicolumn{2}{c||}{LibriSpeech} & \multicolumn{2}{c|}{Loquacious} \\
        \cline{4-7}
            & & & \makecell{dev- \\ other} & \makecell{test- \\ other} & dev & test \\
        \hline\hline
        \multirow{3}{*}{\makecell{From \\ scratch}}
        & AED & - & 4.21 & \textbf{4.37} & 5.35 & 5.93 \\
        \cline{2-7}
        & Prefix LM & - & 4.28 & 4.46 & \multicolumn{2}{c|}{n.a.} \\
        \cline{2-7}
        & Merged attention & - & 4.20 & 4.56 & \multicolumn{2}{c|}{n.a.} \\
        \hline\hline
        \multirow{4}{*}{\makecell{Fine- \\ tuning}} & \multirow{2}{*}{Prefix LLM} & 0.5B & 4.37 & 4.63 & 5.41 & 5.94 \\
        \cline{3-7}
         & & 1.5B & \textbf{4.15} & 4.5 & \textbf{5.28} & \textbf{5.80} \\
        \cline{2-7}
         & \multirow{2}{*}{Merged attention} & 0.5B & 4.35 & 4.67 & 5.49 & 5.97 \\
         \cline{3-7}
         & & 1.5B & 4.33 & 4.67 & 5.39 & 5.85 \\
        \hline
    \end{tabular}
\end{table}


We compare training models with different attention interfaces from scratch
in \Cref{tab:wers-lbs-attention-interfaces}.
We observe similar performance across different attention interfaces,
with the AED model performing slightly better than the PLM and the merged attention model.

In the same table,
we also report results when fine-tuning from the baseline AED encoder and a Qwen2 decoder.
In our initial experiments, we were not able to successfully
fine-tune an AED model from a pre-trained LLM,
which is why we omit these results.
As with the from-scratch models, the fine-tuned PLLM and merged attention models
perform similarly, with the merged attention model performing slightly worse than the PLLM.
Our PLLM results are roughly on-par with the from-scratch AED results on Loquacious,
even though the PLLM is initialized with a pre-trained LLM which is both larger
than the AED decoder and is pre-trained on huge amounts of text data.
To rule out the possibility of undertraining, we try fine-tuning the 0.5B PLLM for
up to 16 epochs on Loquacious. 
However, both for using the 10K SPM ASR vocabulary and the 150K BPE LLM vocabulary, 
we are not able to improve over the from-scratch AED results
when training for the same number of epochs.

\subsection{Combining Multiple Prefix LLM Scores}

\begin{table}
	\centering
	\caption{
        Effect of \textbf{combining multiple prefix LLM (PLLM) scores}
        on Loquacious dev.
        By forwarding the label context with and without the encoder 
        output through the decoder, we can use the PLLM both
        as an acoustic model and as a language model and combine the scores.
        All models are initialized with the baseline AED encoder 
        and Qwen2 LLM.
        We fine-tune for 2 epochs on Loquacious.
        Models with ASR vocabulary are fully fine-tuned, 
        while models with LLM vocabulary use LoRA for the decoder.
        Results using label-synchronous beam search with different 
        combinations of PLLM, CTC, and LLM* scores.
        LLM* refers to using the PLLM decoder without prefixing the encoder output.
    }
	\label{tab:integrated-model-comb-wers-loq}
	\setlength{\tabcolsep}{3pt}
    \begin{adjustbox}{max width=\linewidth}
    \begin{tabular}{|c|c|c|c|c|c|c|c|c|c|}
        \hline
        \multirow{4}{*}{\makecell{LLM \\ size}} & \multirow{4}{*}{\makecell{Label \\ units}} & \multicolumn{8}{c|}{WER [\%]} \\
        \cline{3-10}
        & & \multicolumn{4}{c|}{dev} & \multicolumn{4}{c|}{test} \\
        \cline{3-10}
        & & \scriptsize PLLM & \scriptsize \makecell{PLLM \\ +CTC} & \scriptsize \makecell{LLM* \\ +CTC} & \scriptsize \makecell{PLLM \\ +LLM* \\ +CTC\phantom{0}} & \scriptsize PLLM & \scriptsize \makecell{PLLM \\ +CTC} & \scriptsize \makecell{LLM* \\ +CTC} & \scriptsize \makecell{PLLM \\ +LLM* \\ +CTC\phantom{0}} \\
        \hline\hline
        \multirow{2}{*}{0.5B} & \asrvocab & 5.76 & 5.41 & 5.78 & 5.40 & 6.64 & 5.94 & 6.32 & 5.91 \\
        \cline{2-10}
         & \llmvocab & 6.05 & 5.79 & 5.89 & 5.66 & 6.75 & 6.38 & 6.47 & 6.25 \\
        \hline
        \multirow{2}{*}{1.5B} & \asrvocab & \textbf{5.47} & \textbf{5.28} & \textbf{5.63} & \textbf{5.22} & \textbf{6.03} & \textbf{5.80} & \textbf{6.23} & \textbf{5.72} \\
        \cline{2-10}
         & \multirow{2}{*}{\llmvocab} & 5.78 & 5.67 & 5.72 & 5.46 & 6.36 & 6.26 & 6.33 & 6.03 \\
        \cline{1-1} \cline{3-10}
        7B &  & 5.54 & 5.52 & 5.76 & 5.28 & 6.13 & 6.05 & 6.41 & 5.89 \\
        \hline
    \end{tabular}
    \end{adjustbox}
\end{table}

PLLMs can be used as standalone LMs by not feeding the encoder output into the decoder, 
or as acoustic models by feeding the encoder output into the decoder.
We can also use them in a combined way,
i.e. by forwarding the label context with and without the encoder 
output through the decoder and then combining the scores.
We investigate the effect of using PLLMs in such ways in \Cref{tab:integrated-model-comb-wers-loq}.
Across different LLM sizes, we find that using the PLLM both as an acoustic model 
and as a language model, together with the CTC scores, leads to the best performance.
We note that we never train our PLLMs by just feeding the label context without the encoder output.

\subsection{Downsampling}
\label{subsec:downsampling}

\begin{table}
	\centering
	\caption{
        Effect of \textbf{different downsampling strategies in the adapter} of prefix LLMs
        (PLLM) on Loquacious dev.
        All models are initialized with the baseline AED encoder and the Qwen2 0.5B decoder.
        We fine-tune for 1 epoch on Loquacious using LoRA and the original 150K BPE vocabulary for the decoder.
        For CTC compression, we calculate the average downsampling factor on Loquacious dev.
        Memory usage, RTFs, and WERs [\%] on Loquacious dev.
        Memory usage and RTF are shown for joint CTC recognition.
    }
	\label{tab:downsampling-wers-loq}
	\setlength{\tabcolsep}{2pt}
    \begin{adjustbox}{max width=\linewidth}
    \begin{tabular}{|c|c|c|c|c|c|c|c|c|}
        \hline
        \multicolumn{3}{|c|}{Downsampling} & \multicolumn{2}{c|}{\makecell{Max. GPU \\ mem. [GB]}} & \multicolumn{2}{c|}{Time} & \multicolumn{2}{c|}{WER [\%]} \\
        \hline
        Method & \makecell{CTC \\ thresh.} & Factor & Train & Recog & \makecell{Train\\ {[h]}} & \makecell{Rec. RTF \\ {[$10^{-3}$]}} & PLLM & \makecell{PLLM \\ +CTC} \\
        \hline\hline
        \multirow{3}{*}{\makecell{Concat \\ frames}} & \multirow{3}{*}{-} & 1 & 85.8 & 9.4 & 32.1 & 6.9 & \textbf{6.06} & 5.99 \\
        \cline{3-9}
         & & 2 & 82.4 & 9.3 & 29.7 & 6.8 & 6.21 & 5.99 \\
        \cline{3-9}
         & & 20 & \textbf{74.9} & \textbf{9.2} & \textbf{28.1} & \textbf{6.6} & 6.95 & 6.11 \\
        \hline
        \multirow{3}{*}{\makecell{CTC \\ comp.}} & 0.9 & $\sim$ 3.5 & 89.3 & 9.4 & 29.6 & 6.8 & 6.07 & \textbf{5.98} \\
        \cline{2-9}
         & 0.6 & $\sim$ 3.6 & 85.5 & 9.3 & 29.2 & 6.8 & 6.07 & 5.95 \\
        \cline{2-9}
         & 0.3 & $\sim$ 3.7 & 85.2 & 9.3 & 29.8 & 6.9 & 6.14 & 6.01 \\
        \hline
    \end{tabular}
    \end{adjustbox}
\end{table}

We compare CTC compression to static downsampling for the adapter in \Cref{tab:downsampling-wers-loq}.
As expected, memory usage, training time, and recognition time are reduced with higher downsampling factors.
While the performance of the standalone PLLM degrades with higher static downsampling factors,
it stays on-par with no downsampling when using CTC compression.
Interestingly, the performance with joint CTC recognition remains relatively stable, even when 
using a static downsampling factor of 20.

\subsection{Fine-Tuning Settings}

\begin{table}



	\centering
	\caption{
        Effect of different \textbf{decoder fine-tuning strategies and training
        schedules} for prefix LLMs (PLLM).
        All models are initialized with the baseline AED encoder and the Qwen2 0.5B
        decoder and fine-tuned for 1 epoch on Loquacious using the original 150K BPE vocabulary.
        For the training schedules, \textit{one-stage} refers to directly fine-tuning
        all parameters from the start, while \textit{two-stage} refers to first training
        only the newly initialized parameters and then fine-tuning all weights jointly.
    }
	\label{tab:ft-strategies-wers-loq}
	\setlength{\tabcolsep}{3pt}
    \begin{adjustbox}{max width=\linewidth}
    \begin{tabular}{|c|c|c|c|c|c|c|}
        \hline
        \multirow{3}{*}{\makecell{Decoder \\ FT}} & \multirow{3}{*}{Schedule} & \multirow{3}{*}{\makecell{\# Trainable \\ params. [B]}} & \multicolumn{4}{c|}{WER [\%]} \\
        \cline{4-7}
        & & & \multicolumn{2}{c|}{PLLM} & \multicolumn{2}{c|}{\makecell{PLLM \\ +CTC}} \\
        \cline{4-7}
            & & & dev & test & dev & test \\
        \hline\hline
        Full & \multirow{2}{*}{One-stage} & 1.1 & \textbf{6.06} & \textbf{6.69} & \textbf{5.72} & \textbf{6.26} \\
        \cline{1-1} \cline{3-7}
        \multirow{2}{*}{LoRA} & & \multirow{2}{*}{0.6} & 6.21 & 6.80 & 5.99 & 6.57 \\
        \cline{2-2} \cline{4-7}
        & Two-stage & & 6.20 & 6.88 & 5.75 & 6.27 \\
        \hline
    \end{tabular}
    \end{adjustbox}
\end{table}

We show the effect of different fine-tuning variants and schedules in 
\Cref{tab:ft-strategies-wers-loq}.
We find that full fine-tuning of the decoder performs better than using LoRA,
both with and without joint CTC recognition. In case of LoRA, we further tested
a two-stage training schedule, where we first train only the newly initialized weights 
and then fine-tune all trainable parameters jointly. While performing similarly
in the standalone PLLM case, this two-stage training schedule improves 
performance in the joint CTC recognition case.



\subsection{Final Results}

\begin{table}
	\centering
	\caption{
        Results on the \textbf{HuggingFace ASR leaderboard}.
        We use the official text normalization on the model output
        and reference transcript for evaluation.
        Our models are trained on Loquacious.
        The baseline AED model is trained from scratch for
        20 epochs, while the prefix LLM (PLLM) is initialized with the 
        baseline AED encoder and the Qwen2 0.5B decoder
        and fine-tuned for 2 epochs on Loquacious using LoRA for the decoder.
        Shallow fusion (CTC+LLM) uses the baseline CTC model and our 
        fine-tuned Qwen2 0.5B decoder.
        Comparing to other models from the literature and leaderboard.
    }
	\label{tab:asr-leaderboard-wers}
	\setlength{\tabcolsep}{2pt}
    \begin{adjustbox}{max width=\linewidth}
    \begin{tabular}{|c|c|c|c|c|c|c|c|c|c|}
        \hline
        \multirow{3}{*}{System} & \multicolumn{9}{c|}{WER [\%]} \\
        \cline{2-10}
            & \multirow{2}{*}{\scriptsize AMI} & \multirow{2}{*}{\scriptsize \makecell{Earn\\ings22}} &  \multirow{2}{*}{\scriptsize \makecell{Giga\\speech}} & \multicolumn{2}{c|}{\scriptsize LBS} & \multirow{2}{*}{\scriptsize \makecell{SPGI\\Speech}} & \multirow{2}{*}{\scriptsize \makecell{Ted\\lium}} & \multirow{2}{*}{\scriptsize \makecell{Vox\\populi}} & \multirow{2}{*}{\scriptsize Avg.} \\
        \cline{5-6}
            &  &  &  & \scriptsize Clean & \scriptsize Other &  &  &  & \\
        \hline\hline
        \multicolumn{10}{|c|}{Others} \\
        \hline
        \scriptsize \makecell{Canary-Qwen-2.5B \\ (Qwen3 1.7B)} & \textbf{10.19} & \textbf{10.45} & 9.43 & 1.61 & 3.10 & \textbf{1.90} & \textbf{2.71} & \textbf{5.66} & \textbf{5.63} \\
        \hline
        \scriptsize \makecell{Qwen3-ASR-0.6B \\ (Qwen3 0.6B)} & 11.66 & 11.06 & \textbf{9.14} & 2.13 & 4.45 & 3.03 & 2.85 & 7.07 & 6.42 \\
        \hline\hline
        \multicolumn{10}{|c|}{Ours} \\
        \hline
        \makecell{CTC + LLM \\ (Qwen2 0.5B)} & 21.11 & 34.19 & 12.95 & \textbf{1.52} & \textbf{2.99} & 5.62 & 4.11 & 6.00 & 11.06 \\
        \hline
        \makecell{CTC + PLLM \\ (Qwen2 0.5B)} & 20.06 & 30.01 & 11.59 & 1.95 & 3.79 & 4.45 & 4.24 & 6.23 & 10.29 \\
        \hline
    \end{tabular}
    \end{adjustbox}
\end{table}


We compare our best results to other models from the literature on the HuggingFace ASR leaderboard in 
\Cref{tab:asr-leaderboard-wers}.
We use the official text normalization%
\footnote{\scriptsize\url{https://github.com/huggingface/open_asr_leaderboard/blob/main/normalizer/normalizer.py}} 
on the model output and reference transcript for evaluation.
We observe that, while shallow fusion outperforms our tightly integrated PLLM on Loquacious and LibriSpeech
(\Cref{tab:tight-vs-shallow-wers-loq}), the opposite is the case on the HuggingFace ASR leaderboard, where our PLLM outperforms shallow fusion.
More specifically, shallow fusion performs better on the in-domain corpora (LibriSpeech and Voxpopuli),
which are also part of Loquacious, as well as Tedlium, while our PLLM performs better on the out-of-domain corpora 
(AMI, Earnings22, Gigaspeech, and SPGI Speech).
Further analysis is needed to better understand the reasons for this behavior.
Compared to the other models on the leaderboard, our systems are drastically worse on the AMI and Earnings22 corpora
-- far-field meeting recordings and accented earnings-call speech, respectively --
while being closer, and sometimes even better, on the other corpora.
We note that the other models on the leaderboard are trained on magnitudes more ASR data than our models, which makes the comparison difficult.

\section{Conclusions}

In this work, we have investigated different utilizations of
large language models (LLMs) for automatic speech recognition (ASR).
We compared tight integration of acoustic encoder and
LLMs, i.e. \textit{speech LLMs} (SLLMs),
to shallow fusion of acoustic and language model scores.
Our findings are as follows:

\begin{itemize}
    \item On in-domain evaluations, shallow fusion of LLM 
    and CTC scores usually outperforms or matches tight integration 
    of acoustic encoder and LLM. This highlights the importance of 
    comparing tightly integrated speech LLMs against strong 
    shallow-fusion baselines.
    \item On Loquacious and LibriSpeech, under comparable model sizes,
    our SLLM performance is on par with our
    AED baseline which was solely trained on the ASR data.
    This questions the benefit of using pre-trained LLMs for ASR
    compared to specialized ASR models.
    \item Joint CTC+SLLM decoding can significantly improve the 
    performance of standalone SLLM decoding, 
    especially for smaller LLMs.
    For this, we presented two optimizations
    without which the computational cost of joint decoding 
    becomes prohibitive.
    When moving to larger LLMs
    the performance of standalone SLLM decoding and joint
    decoding converges.
    \item SLLMs can be used as both acoustic and language
    models, depending on whether the encoder output is used
    or not. We find that combining these scores together
    with CTC yields improvements over
    CTC+SLLM decoding.
    \item Fine-tuning SLLMs with a smaller vocabulary, which was fitted on the 
    transcription data, improves ASR performance compared
    to using the original vocabulary of the LLM.
    \item The prefix (PLLM) interface slightly outperforms the
    merged attention interface when starting from a pre-trained
    LLM decoder, while the two perform similarly when trained
    from scratch.
\end{itemize}




\section{Generative AI Use Disclosure}
We use LLMs to improve the formulations and grammar of the paper.

\ifbool{blind}{
  
}{
\section*{Acknowledgment}

{
    \footnotesize
    This work was partially supported by NeuroSys, which as part of the initiative “Clusters4Future” is 
    funded by the Federal Ministry of Research, Technology and Space BMFTR (funding IDs 03ZU2106DA and 03ZU2106DD).
    The authors gratefully acknowledge the computing time provided to 
    them at the NHR Center NHR4CES at RWTH Aachen University 
    (project number p0023999).
    This is funded by the Federal Ministry of Education and Research, 
    and the state governments participating on the basis of the resolutions 
    of the GWK for national high performance computing at universities 
    (\url{www.nhr-verein.de/unsere-partner}).
    }
}

\bibliographystyle{IEEEtran}
\bibliography{mybib}

\end{document}